\documentclass[pdftex]{article}
\usepackage{icrctc07}

\title{Observations of the high-frequency-peaked BL Lac object 1ES 1218+304 with VERITAS}
\shorttitle{Observations of the high-frequency-peaked BL Lac object 1ES 1218+304 with VERITAS}

\authors{Pascal Fortin$^{1}$ for the VERITAS collaboration$^{2}$}
\shortauthors{Fortin and et al.}
\afiliations{$^1$Dept. of Physics and Astronomy, Barnard College, Columbia University, New York, NY, 10027 \\ $^2$For full author list see G. Maier, "VERITAS: Status and Latest Results", these proceedings}
\email{fortin@phys.columbia.edu}

\abstract{The VERITAS collaboration has observed the high-frequency-peaked BL Lac object 1ES 1218+304 using an array of several imaging Cherenkov telescopes located at the Fred Lawrence Whipple Observatory in Southern Arizona. A gamma-ray signal was detected with high significance for the observations taken during several months in the 2006-2007 observing season. Here we present the detection of 1ES 1218+304 in very-high-energy gamma rays.}

\begin{document}
\maketitle
\section{Introduction}

One of the major discoveries from the Energetic Gamma Ray Experiment Telescope (EGRET) was the detection of high-energy emission from more than 60 Active Galactic Nuclei (AGN) of the blazar class \cite{hartman}. Blazars, which include BL Lac objects and flat-spectrum radio quasars (FSRQs), are characterized by non-thermal emission and their spectral energy distribution (SED) contains two broad peaks. The low-energy peak (radio to UV or X-rays) is commonly interpreted as synchrotron radiation from ultra-relativistic electrons moving along a plasma jet pointing at us. The origin of the second peak is less certain. Several models, from pure leptonic or hadronic models to leptonic/hadronic hybrid models, can explain the high-energy peak (X-rays to TeV $\gamma$-rays) \cite{Bottcher:2007fr}. Broadband observations from the radio to Very-High-Energy (VHE) $\gamma$-rays are necessary to understand the physics of the jets and emission mechanisms. VHE observations can also help constrain the intensity and spectrum of the extragalactic background light (EBL), which are important parameters for cosmologists to test our understanding of structure and star formation in the Universe. The EBL at near-IR to mid-IR wavelengths is the dominant absorber of TeV $\gamma$-rays via pair production. The measurement of high-quality energy spectra for blazars over the energy range from 100 GeV to 10 TeV can be used to gain information about the EBL \cite{Dwek:2005ul}.
The number of detected blazars at TeV energies remains small at less than two dozen. With the exception of the recently discovered low-frequency-peaked BL Lac (LBL) object BL Lacertae \cite{Albert:2007mz}, all TeV blazars are high-frequency-peaked BL Lac objects (HBL).

During the first two years of operation of VERITAS, four key science projects will jointly receive 50\% of the observing time. Observations of blazars is one of them. One of the objects selected for the blazar key science project is the HBL object 1ES 1218+304, an X-ray-bright ($F_{1 keV}>2\mu $Jy), ``extreme" BL Lac located at a redshift $z=0.182$ \cite{Bade:1998ly}. Based on SED modeling and \textit{Beppo}SAX X-ray spectra, several of the ``extreme" HBL were predicted to be TeV sources and several of them have indeed been detected at TeV energies \cite{Costamante:2002rt}. 1ES 1218+304 was predicted to be a good TeV candidate based on the position of its synchrotron peak at high energy and sufficient radio-to-optical flux. It was recently detected (at very high energy) by the MAGIC telescope \cite{Albert:2006yq}. Here we report on the detection of 1ES 1218+304 in VHE gamma rays with VERITAS.

\section{Observations}

The VERITAS observatory uses an array of four 12-meter imaging atmospheric Cherenkov telescopes (IACTs) located at the F. L. Whipple Observatory ($31^\circ 40'30.21''$ N, $110^\circ57'07.77''$ W, 1268 m a.s.l) in southern Arizona \cite{T.-C.-Weekes:2002lr}. The telescopes use the Davies-Cotton design and are made of 345 front-aluminized and anodized hexagonal glass facets \cite{Roache:2007fk}. Each camera consists of 499 photomultiplier tubes (PMTs) separated by $0.15^\circ$ and covers a $3.5^\circ$ field of view. Light concentrators reduce the dead-space between PMTs and decrease the amount of night-sky light seen by the PMTs \cite{Nagai:2007qy}. The analog signals from the PMTs are pre-amplified in the camera before being sent through coaxial cables to an electronics trailer located at the base of each telescope.

VERITAS uses a three-level trigger system to reduce the rate of background events caused by fluctuations in the night sky light and cosmic-ray showers while retaining multi-telescope images consistent with gamma-ray showers \cite{Weinstein:2007uq}. Upon triggering, the analog PMT signals are digitized using custom-designed 500 Mega-Sample-Per-Second flash-analogue-to-digital converters (FADCs) and the data are archived to disk. Additional details about the data acquisition system can be found in \cite{Hayes:2007qy}.

The first two telescopes were operated in array mode from March 2006 and the third and forth telescopes came online in December 2006 and April 2007 respectively \cite{Maier:2007lr}. VERITAS observed 1ES 1218+304 from December 2006 to March 2007 using two or three telescopes. A small fraction of the data was taken in pair mode where the source is observed for 28 minutes (ON), followed 2 minutes later by the same exposure offset in right ascension by 30 minutes (OFF).
The bulk of the data were taken in \textit{wobble} mode where the source is offset from the center of the field of view by $0.3^\circ-0.5^\circ$ and the background is measured directly from regions which are at identical offsets from the center of the field of view but away from the source region. Here we report results for observations taken only in \textit{wobble} mode at an offset of $0.5^\circ$ with three telescopes. After removing data taken under poor sky conditions or affected by various detector problems, we are left with a total observation time of 17.4 hours covering a range in zenith angle from $2^{\circ}$ to $35^{\circ}$.

\section{Data Analysis \& Results}

The analysis of the data was made using independent analysis packages (see \cite{Daniel:2007lr,Cogan:2007lr} for details on the analyses). All of these analyses yield consistent results.
After cleaning the images by cutting on the integrated charge, the standard Hillas parameters \cite{hillas85} are calculated. The location of the shower source in the field of view is calculated by minimizing the weighted perpendicular distance to each image axis.
A set of \textit{scaled cuts} on the \textit{width} (MSW) and \textit{length} (MSL) parameters are used to identify gamma-ray events in the data. The cuts were optimized using data from the Crab Nebula \cite{Celik:2007fj}. An extensive set of Monte Carlo simulations \cite{Maier:2007kx} was used to determine the effective area of the detector as a function of zenith angle and gamma-ray energy. Lookup tables were generated from these simulations to calculate the energy of the primary gamma rays.

\begin{figure}
\begin{center}
\includegraphics [width=0.45\textwidth]{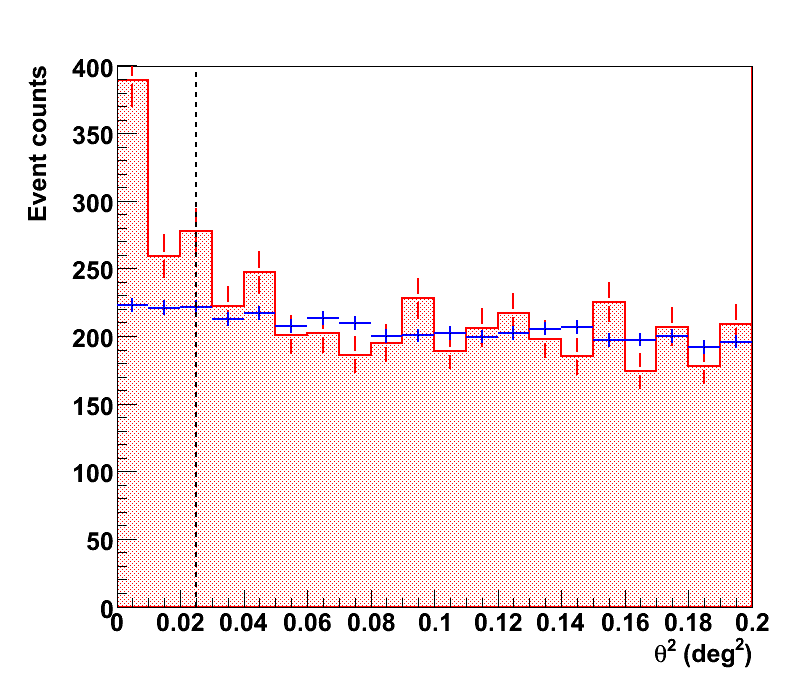}
\end{center}
\caption{$\theta^2$ distribution for \textit{signal} (solid histogram) and \textit{background} (dots) regions. The dashed line indicates the boundary of the \textit{signal} region at $0.158^\circ$ as determined from optimization on the Crab Nebula.}\label{fig1}
\end{figure}

Figure \ref{fig1} shows the distribution of the squared angular distance ($\theta^2$) between the reconstructed shower directions and 1ES1218+304 coordinates (solid histogram), and between the shower directions and the \textit{background} regions coordinates (dots). A clear excess is visible below our angular cut of $0.158^\circ$, corresponding to a detection significance of 8.9 standard deviations.

\begin{figure}
\begin{center}
\includegraphics [width=0.45\textwidth]{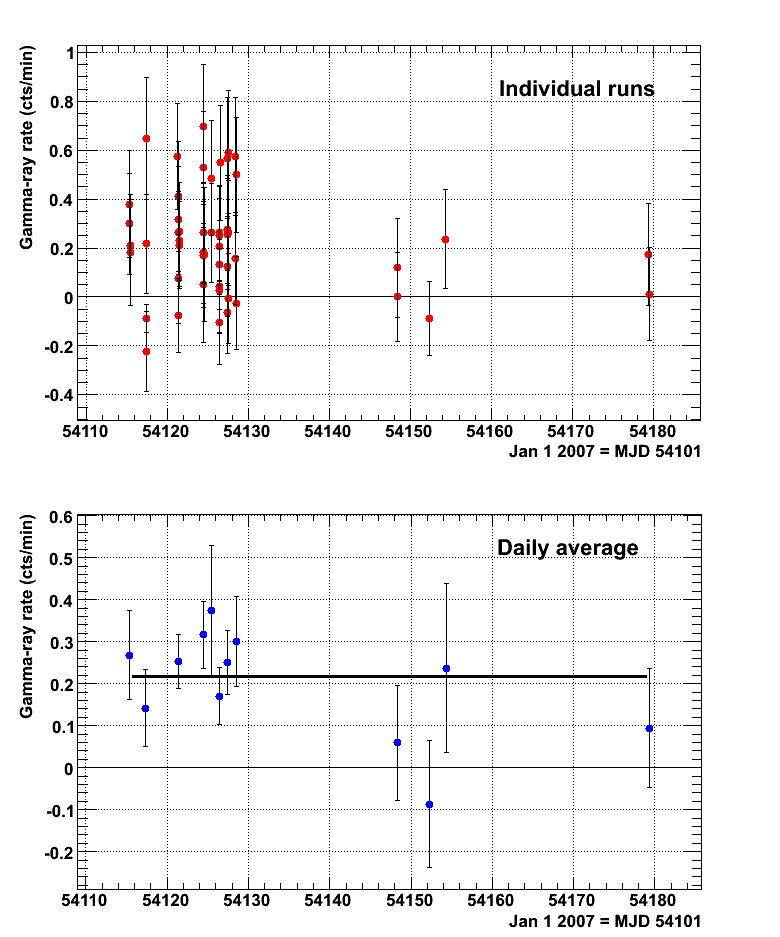}
\end{center}
\caption{Excess count rate (min$^{-1}$) for each run (top panel) and daily average (lower panel) for the source 1ES 1218+304. The count rates were not corrected for zenith-angle dependence. Error bars represent the statistical uncertainty only.}\label{fig2}
\end{figure}

Figure \ref{fig2} shows the light curve for the months of January, February and March. The top panel shows a data point for each 20-minute run and the lower panel shows the daily averages. These results are considered preliminary as the excess rates were not corrected for zenith-angle dependence.

\section{Discussion and conclusions}

The results presented here confirm with high statistical significance the MAGIC discovery \cite{Albert:2006yq} of the HBL object 1ES 1218+304 as a source of VHE gamma rays.

\subsection*{Acknowledgments}

This research is supported by grants from the U.S. Department of Energy, the U.S. National Science Foundation, and the Smithsonian Institution, by NSERC in Canada, by PPARC in the UK and Science Foundation Ireland.

\bibliography{icrc0564}
\bibliographystyle{unsrt}
\end{document}